\documentclass[aps,prl,twocolumn,nofootinbib,groupedaddress]{revtex4}

\usepackage{graphicx}
\begin{document}

\title{Gravitational Wave Production At The End Of Inflation}

\author{Richard Easther}
\author{John T. Giblin, Jr}
\author{Eugene A. Lim}

\affiliation{Dept. of Physics, Yale University, New Haven CT 06520 }
\date{\today}

\begin{abstract}
We consider gravitational wave production due to parametric resonance at the end of inflation, or  ``preheating''.   This leads to large inhomogeneities which source a stochastic background of gravitational waves at scales  inside the comoving Hubble horizon at the end of inflation.  We confirm that the present amplitude of these gravitational waves  need not depend on the inflationary energy scale. We analyze an explicit model where the inflationary energy scale is $\sim 10^9$ GeV, yielding a signal close to the sensitivity of  Advanced LIGO and BBO. This signal highlights the possibility of a new observational ``window'' into inflationary physics, and provides significant motivation for searches for stochastic backgrounds of gravitational waves in the  Hz to GHz range, with an amplitude on the order of $\Omega_{gw}(k)h^2 \sim 10^{-11}$.  Finally,  the strategy used in our numerical computations is applicable to the gravitational waves generated by many inhomogeneous  processes in the early universe.
\end{abstract}


\maketitle

 
A successful model of inflation must have  a ``graceful exit" that describes the transition from the accelerated phase to a thermalized universe  \cite{Guth:1980zm}.   A widely studied mechanism for
achieving this is preheating (e.g.
\cite{Traschen:1990sw,Kofman:1994rk,Garcia-Bellido:1997wm,Khlebnikov:1997di,Greene:1997ge,Parry:1998pn,Bassett:1998wg,Garcia-Bellido:1998gm,Easther:1999ws,Liddle:1999hq,Finelli:2001db,Bassett:2005xm,Podolsky:2005bw,Easther:2006gt,Felder:2006cc}).
After inflation, the inflaton (or a related field) oscillates about the bottom
of its potential, driving the resonant amplification of specific momentum
modes of some coupled field(s).    This renders the universe inhomogeneous, and the resulting spatial gradients
source gravitational waves. For
GUT inflation, the present peak frequency is between 1 MHz and 1 GHz \cite{Khlebnikov:1997di,Easther:2006gt}.  It was conjectured that the characteristic frequency is inversely proportional to the inflationary scale, while the amplitude can be  {\em independent\/} of this scale \cite{Easther:2006gt}, leading to a signal potentially detectable by future iterations of LIGO and BBO. We confirm this conjecture by  numerically computing the gravitational wave spectrum in a toy model of preheating following low scale inflation.   While simple inflationary models typically involve GUT scale physics, many stringy models have a much lower inflationary scale, so this signal may eventually lead to new constraints on these models.   The tools developed for this analysis will allow us to explore fully realistic preheating models  in an expanding background. Furthermore, our computational strategy applies to any inhomogeneous phase in the universe, and may have applications beyond the present problem.

{\bf  Computational Strategy \& Results:}  During parametric
resonance, momentum modes of a field $\chi$ are pumped by an oscillating field
$\phi$.  In simple models $\phi$ is the inflaton, but in hybrid models $\phi$
is the direction orthogonal to the inflationary trajectory which induces the
``waterfall'' transition  \cite{Linde:1993cn,Copeland:1994vg}.  In either case
the lagrangian can be expressed as  
\begin{equation}
\mathcal{L} = -\frac{1}{2}\left(\partial\phi \right)^2 -
\frac{1}{2}\left(\partial\chi\right)^2  - V(\phi,\chi). 
\end{equation}
We numerically simulate the nonlinear field evolution in a conformally rigid spacetime background.  We can then compute the spatial parts of $T_{\mu\nu}$ at any given time. The tensor  contribution to the metric perturbation and Einstein equations read
\begin{eqnarray}
&ds^2 = dt^2 - a^2(t)\left(\delta_{ij} + h_{ij}\right)dx^idx^j & \, ,\\
&\bar{G}_{\mu\nu}(t) + \delta G_{\mu\nu}(x,t) = {8\pi G} \left[ 
\bar{T}_{\mu\nu}(t) + \delta T_{\mu\nu}(x,t) \right]&
\end{eqnarray}
where the overbar denotes the homogeneous background values.  The perturbation is {\it Transverse-Traceless}, so
\begin{equation}
h^i_i = 0\, \qquad  h^i_{j,i} = 0 \, .
\end{equation}
%
%
The Fourier components of the $h_{ij}$ obey
\begin{equation}
\ddot{\tilde{h}}_{ij}  +
3\frac{\dot{a}}{a}\dot{\tilde{h}}_{ij}  - 
2\left(\frac{\dot{a}^2}{a^2} - 2\frac{\ddot{a}}{a} -
\frac{k^2}{a^2} \right)\tilde{h}_{ij} =\frac{ 16\pi G}{a^2} \tilde{S}^{TT}_{ij}  \label{source}
\end{equation}
using the convention
\begin{equation}
\tilde{f}(k,t) = \int d^3 x f(x,t) e^{2\pi i k\cdot x} \, 
\end{equation}
and keeping only the transverse-traceless source. 

\begin{figure*}[htb]
\includegraphics[width=7in]{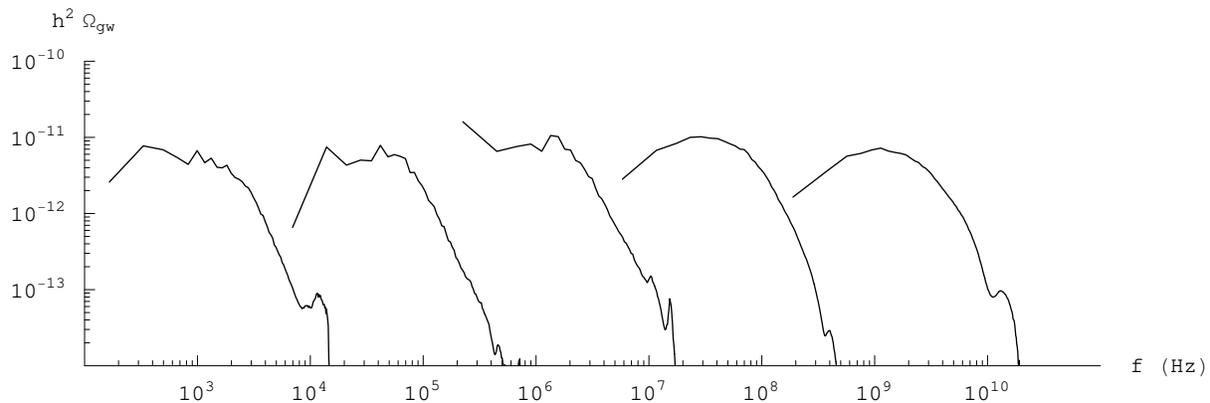}   
\caption{We plot the spectrum of gravitational radiation produced during   resonance with   $\mu =  10^{-18}$ (left) through to $10^{-6}$  (right) in units where $m_{Pl} \approx  10^{19} GeV =1$, where each spectrum has a value of $\mu$ $10^3$ times large than the one immediately to the left.  The corresponding initial energy densities run from  from ($4.5 \times10^{9} {\rm GeV})^4$ to ($4.5 \times10^{15} {\rm GeV})^4$ for our choice of $\phi_0$.  The plots are made on $128^3$ grids, and the ``feature" at high frequency is a numerical artifact.  \label{fig:comboresults}}
\end{figure*}

We evolve the fields with {\sc LatticeEasy} \cite{Felder:2000hq},  treating the spatial background as a rigid, expanding box.  We assume the fields' initial inhomogeneity is derived from their quantum mechanical vacuum fluctuations.   The $\tilde{h}_{ij}$ obey {\em ordinary} differential equations, which we solve with a fourth order Runge-Kutta integrator.  Unlike   \cite{Khlebnikov:1997di,Easther:2006gt}, we use no flat-space results, and the gravitational wave amplitude can be followed throughout the simulations.   Numerical noise is a significant problem, since the transverse-traceless source involves differences of terms involving derivatives, which are themselves computed via numerical differencing on the lattice -- which can easily lead to a loss of numerical significance.  In future work we plan to perform simulations at higher spatial resolution, with different algorithms for the field evolution, in order to produce highly accurate spectra for a variety of models over a large frequency range. Moreover, we have made simplifying assumptions about the subsequent evolution of the universe, which can rescale the overall tensor spectrum.  In models without resonance, the computed tensor spectrum is many orders of magnitude smaller than seen in resonant models, providing a ``null check'' of our algorithm, and we have recovered the key results of \cite{Khlebnikov:1997di,Easther:2006gt}. Compared to these results, the spectra in \cite{Khlebnikov:1997di,Easther:2006gt} are noticeably noisy, which is an artifact of the  older algorithm. The amplitudes obtained with our new code are somewhat lower than the previous results, possibly due to better noise suppression.
 
We calculate the power spectrum using \cite{Chongchitnan:2006pe}
\begin{equation} \label{rho}
\rho_{gw} = \left. \frac{1}{32\pi G} \left<h_{ij,\mu}h^{ij,\nu}\right>  \right|_{\mu = \nu = 0}\, .
\end{equation}
Evolving the $h_{ij}$ and inserting  
(\ref{rho})    into equation (20) of \cite{Easther:2006gt} (and  correcting a typographical error) we find:
\begin{equation} \label{omeganow}
\Omega_{gw}(k) h^2 = \Omega_r h^2 \left(\frac{g_0}{g_\star}\right)^{1/3} 
 \frac{d}{d\ln k}\left[\frac{\rho_{gw}(a_e)}{\rho_{\mbox{\tiny tot}}(a_e)}\right]
\end{equation}
Here $a_e$ is the scale factor at the end of the simulation, $\Omega_r$ is the present-day density of radiation, $g_0$ and $g_\star$  denote the number of thermal degrees of freedom at present and at matter-radiation equality, and $h$ is the dimensionless Hubble constant at the present epoch.  We  thus assume that the universe is radiation dominated between the end of our simulation and matter-radiation equality, during which time $\Omega_{gw}(k)$ is constant, and matter dominated thereafter.   In the future we plan to consider more realistic transfer functions \cite{Boyle:2005se,Watanabe:2006qe}. Moreover, we implicitly assume that the universe promptly thermalizes after preheating, which requires a trilinear coupling we have not included here \cite{Podolsky:2005bw}. 

We   work with 
\begin{equation} \label{Veff}
V(\phi,\chi) = \frac{1}{2}\mu^2\phi^2 +  \frac{1}{2}g^2\phi^2\chi^2  \, .
\end{equation}
In standard quadratic chaotic inflation, $\mu$ is fixed by the power spectrum, and is not a free parameter.  Since we are interested in the end of inflation, equation~(\ref{Veff}) need not describe the potential in the epoch when cosmological perturbations were generated, and   $\mu$ is a free parameter.  To see this, consider 
\begin{equation}
V = \frac{(M^2-\lambda\sigma^2)^2}{4\lambda} +
\frac{m^2}{2}\phi^2 + \frac{h^2}{2} \phi^2\sigma^2 +  \frac{g^2}{2} \phi^2\chi^2  .
\end{equation}
During inflation,  $\phi$ is large  and  $\sigma = 0$.   When $\phi = M/h$,  the field rapidly evolves towards the minimum at $\sigma = M/\sqrt{\lambda}$.  This is the ``waterfall'' phase transition.  Further, we assume that   $\sigma = \langle \sigma \rangle$ and that there is there is no $\sigma$ production via a $\sigma^2\chi^2$ term, giving
\begin{equation} \label{HVeff}
V(\phi) = \frac{1}{2}\left(m^2+\frac{h^2M^2}{\lambda}\right)\phi^2.
\end{equation}
This is a  mass term  for $\phi$.  In practice, we need $m \ll M$ in order to get the correct perturbation spectrum, so the effective mass is simply $\mu^2 = h^2 M^2 / \lambda$.  Here, we will simply take $\mu$ to be a free parameter. The inflationary energy density is then approximately $\mu^2 \phi_0^2$, where $\phi_0$ is the field value at the beginning of our simulation, which we assume to be roughly coincident with the end of inflation. We take $\phi_0\approx.2 m_{pl}$, so the inflationary energy scale is set by  $\mu^{1/2}$.  For hybrid inflation,   $M /h \approx .2 m_{pl}$ and for $\mu =  h M/ \sqrt{\lambda}=10^{-18} m_{pl} \sim 10 \mbox{GeV}$ we need $ {h^4}/\lambda \approx 3 \times 10^{-16}$, which demands a small but not miniscule value of $h$, if we assume $\lambda \sim 0.1$.

We hold the
 resonance parameter $q= g^2 m_{pl}^2 /\mu^2$ fixed \cite{Easther:2006gt}. This ensures that the same modes (in units of post-inflationary energy scale) are in resonance  as $\mu$ is varied.  We choose $q  = 2 \times 10^6$, which makes $g$ unrealistically small as $\mu$ is decreased.  However,  our purpose is to give  an ``existence  proof'' that (p)reheating can generate a substantial gravitational wave background at low inflationary scales. This  assumption ensures that the structure of resonance does not change with the inflationary  energy scale, allowing us to isolate the aspects of gravitational wave production which depend only on the overall inflationary scale.    Figure~\ref{fig:comboresults} shows the gravitational wave spectra for different  inflationary energy scales.  As predicted \cite{Easther:2006gt},    the  peak frequency scales inversely with  energy,  whereas the  amplitude is scale independent.

The spectrum declines very rapidly at large $k$. This reflects the structure of the resonance bands, as modes with large $k$  are never in resonance. Conversely,  we see  a relatively broad plateau, possibly attributable to modes moving through the resonance band as inflation continues.  The amplitude of modes which are super-horizon throughout resonance is expected to  rise as $k^3$ or faster \cite{Liddle:1999hq}.  The universe grows significantly during resonance, so we cannot simultaneously resolve both resonant  and super-horizon modes.  In order to prevent contamination by numerical noise we must choose our box to  ensure that we resolve a significant number of high-frequency modes which are never resonantly amplified.  A very large scale simulation is thus needed to see the full shape of the peak.

The source term is potentially non-zero whenever the fields are inhomogenous, but its amplitude rises rapidly during  resonance, and then decreases as the universe begins to thermalize. Most of the gravitational waves are thus radiated during a comparatively narrow time interval.    In the models treated here, the universe grows $\sim 50\%$ larger as  90\% of the power is generated.   

\begin{figure*}
\includegraphics[width=7.5in]{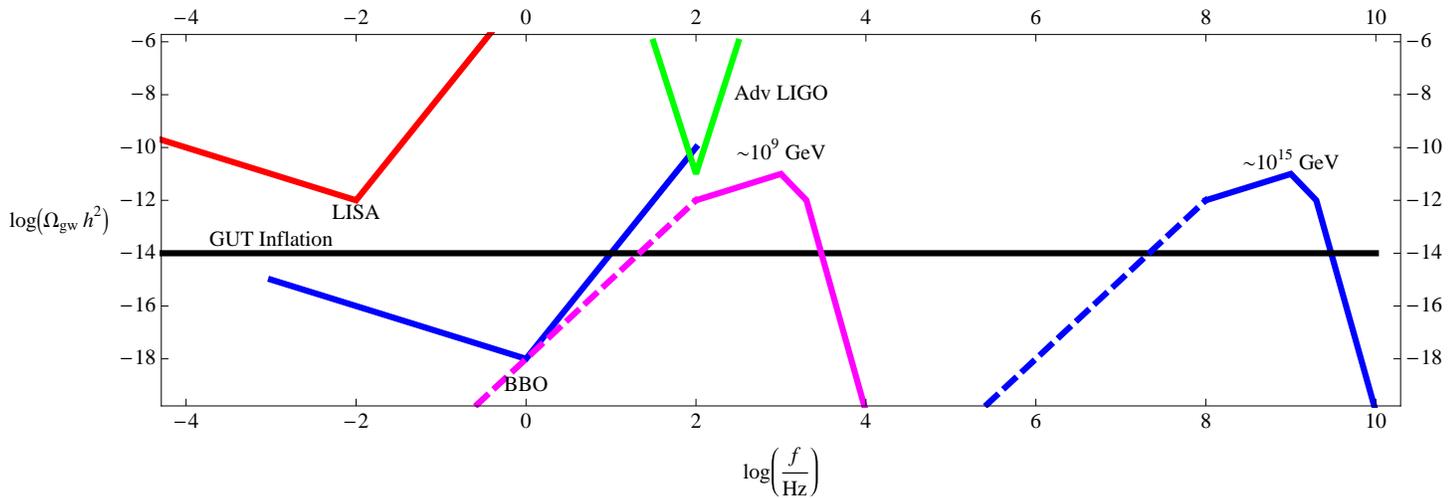}   
\caption{ We sketch the gravitational wave spectra obtained for the lowest and highest energy models computed here, relative to that of the Advanced  LIGO goal, and the proposed LISA and BBO experiments. We see that inflationary models with lower energy scale may lead to a signal which is visible at LIGO scales if the sensitivity of LIGO is further improved, and with BBO. The tensor background generated by quantum fluctuations during GUT scale inflation is shown by the solid horizontal line. The dashed lines denote the inferred $k^3$ tails.  The spectra generated by the inflationary scenarios considered in \cite{Khlebnikov:1997di,Easther:2006gt} roughly overlap with the $10^{15}$ GeV spectrum depicted above.
   \label{fig:sensitivity}}
\end{figure*}

 {\bf Discussion:}  Bubble-wall collisions after a first order phase transition at the end of inflation \cite{La:1989za,Turner:1990rc,Kosowsky:1991ua,Kosowsky:1992rz}  can also generate  a present-day $\Omega_{gw}(k)h^2 \sim 10^{-11}$. This spectrum has similar properties to the preheating background: the frequency scales inversely with the inflationary scale, while the amplitude need  not depend on  the inflationary scale. This process yields   a maximum  $\Omega_{gw}(k)$ similar to that found here for parametric resonance, although the detailed spectra may differ considerably.  This is to be expected since, since gradient terms source ${\tilde{h}}_{ij}$ and the amplitude of these is bounded by  their contribution to the total energy density.  If the gradient energy  saturates this bound,  $\Omega_{gw}(k)h^2 \sim 10^{-11}$ or $10^{-10}$  is a generic result, in the absence of  non-perturbative effects   \cite{Kamionkowski:1993fg,Kosowsky:2001xp}.   In our simulations the gradient energy peaks at around 25\% of the total energy density. Since there are necessarily kinetic and potential contributions to the energy, this fraction can not be significantly increased.

Since our algorithm works for an arbitrary extended source of gravitational radiation, we could  apply it to other mechanisms that generate stochastic backgrounds. These include networks of cosmic superstrings \cite{Siemens:2006yp},  the TeV scale phase transition present in  Randall-Sundrum models \cite{Randall:2006py},  a first order electroweak phase transition  \cite{Grojean:2006bp}, cosmological bubble collisions \cite{Kosowsky:1991ua}, or large-scale  turbulence in the presence of magnetic  fields   \cite{Kamionkowski:1993fg,Kosowsky:2001xp,Kahniashvili:2005qi}.  Moreover,  \cite{Garcia-Bellido:2007dg} presents an alternative (and physically distinct) mechanism for generating gravitational waves at the end of inflation, and we expect that it will be possible to analyze this mechanism with our algorithm.   Finally, \cite{Ananda:2006af} discusses the generation of gravitation waves by  density perturbations during the radiation dominated era, and it would be interesting to explore the overlap between this second order calculation and our techniques.

The detectability of high frequency background of stochastic gravitational radiation is an open question. We plot a schematic version of our results in Figure~\ref{fig:sensitivity}, and we see that the proposed space-based detectors BBO and LISA are sensitive to much longer wavelengths than any signal likely to be generated during preheating from GUT scale inflation.   It will be a stretch for Advanced LIGO to
see the signals computed here.  However, a further iteration of LIGO is likely to put significant constraints on any signal that would be generated during resonance after   inflation occurring at scales around $10^{9}$ GeV.   

Detecting high frequency gravitational waves is particularly challenging, since the required strain-sensitivity at fixed $\Omega_{gw}(k)$ scales as the cube of the frequency. Consequently,  it is easier to detect a background arising from preheating after low-scale inflation. With few exceptions, stringy models of inflation have $V^{1/4} \lesssim 10^{13}$GeV, two orders of magnitude below GUT scales, and can be much lower \cite{Kallosh:2007wm}.   Moreover, reheating after thermal inflation would lead to signal very close to the LIGO band \cite{Felder:2007iz}.  In these models the primordial tensor spectrum is unobservable, but any  preheating signal is easier to detect.    Very low scale inflation (at or near  the TeV scale e.g. \cite{Knox:1992iy,Allahverdi:2006iq}) could produce a signal visible to a  BBO-style mission.   Finally,  future detector technologies might be sensitive to high-frequency gravitational radiation.

Our key result   is an existence proof that parametric resonance following inflation can lead to a significant gravitational wave background, independently of the  inflationary energy scale. Many questions remain. Firstly, we have only considered models which are well-described by (\ref{Veff}), and our set-up is a toy model, requiring a very small coupling parameter at low inflationary scales. This ensures the {\em structure} of resonance is independent of the energy scale, which allows us to isolate the impact on the inflationary energy scale on the height and location of the peak. However, this small coupling is not a prerequisite for preheating or resonance.  Preheating is generically associated with enhanced inhomogeneities which will source gravitational radiation, so we can hope that similar results also apply to more realistic resonance scenarios, but these need to be explored in detail.  Secondly, we have not yet considered overlapping backgrounds from other astrophysical processes. Thirdly, we need to incorporate the finer details of the tensor mode transfer function and  the post-resonance thermalization of the universe.

{\bf Acknowledgments:}
We  thank   Gary Felder, Arthur Kosowsky and David Wands  for useful discussions. We are particularly indebted to Felder and Tkachev for the use of {\sc LatticeEasy\/}.   This work was partially supported   by  the  Department of Energy, DE-FG02-92ER-40704.

\end{document}